\documentclass[onecolumn]{aa}
\usepackage{graphicx}

\begin{document}

\title{Stellar Evolutionary Models for Magellanic Clouds}

\author{V.Castellani \inst{1}$^,$\inst{2}\and
       S. Degl'Innocenti \inst{3}$^,$\inst{4} \and M. Marconi \inst{5} \and P. G. Prada
       Moroni \inst{1}$^,$\inst{2}$^,$\inst{6} \and P. Sestito \inst{7} }

 \offprints{S. Degl'Innocenti}

\institute{ Osservatorio Astronomico di Roma, via Frascati 33, 00040 Monte Porzio
       Catone, Italy
\and
       INFN Sezione di Ferrara, via Paradiso 12,
       44100 Ferrara, Italy
\and
       Dipartimento di Fisica,
       Universit\`a di Pisa, via Buonarroti 2, 56127 Pisa, Italy\\
       \email{scilla@df.unipi.it}
\and
       INFN Sezione di Pisa, via Buonarroti 2, 56127 Pisa, Italy\\
\and
       Osservatorio Astronomico di Capodimonte, via Moiariello 16, 80131 Napoli, Italy
\and
       Osservatorio Astronomico di Collurania, via Maggini, 64100
       Teramo, Italy
\and
       Dipartimento di Astronomia e Scienza
       dello Spazio, Largo E. Fermi 5, 50125 Firenze, Italy }

\date{}

\abstract{
We supplement current evolutionary computations concerning Magellanic
Cloud stars by exploring the evolutionary behavior of canonical
stellar models (i.e.,with inefficient core overshooting) with
metallicities suitable for stars in the Clouds. After discussing the
adequacy of the adopted evolutionary scenario, we present evolutionary
sequences as computed following a selected sample of stellar models in
the mass range 0.8$\div$ 8 M$_{\odot}$ from the Main Sequence till the C
ignition or the onset of thermal pulses in the advanced Asymptotic
Giant Branch phase. On this basis, cluster isochrones covering the
range of ages from $\sim$100 Myr to $\sim$ 15 Gyr are presented and
discussed. To allow a comparison with evolutionary investigations
appeared in the recent literature, we computed additional sets of
models which take into account moderate core overshooting during the H burning
phase, discussing the comparison in terms of current uncertainties in the
stellar evolutionary models. Selected predictions constraining the
cluster ages are finally discussed, presenting a calibration of the
difference in magnitude between the luminous MS termination and the He
burning giants in terms of cluster age. 
Both evolutionary tracks and isochrones have been made
available at the node http://gipsy.cjb.net as a first step of a planned
``Pisa Evolutionary Library''.

}
\maketitle
\section{Introduction}
Stellar evolution in both Magellanic Clouds (MCs) has been the object
of several investigations already appeared in the current literature,
mainly because the stars in the Clouds can provide relevant
constraints on distances to these stellar systems which, in turn, play
a relevant role in constraining the cosmic distance scale (see e.g. 
Gallart et al. 2003, Woo et al. 2003, Bertelli et al. 2003, Salaris et al. 2003).
  At the same time, recent improvements in the observational capabilities have
suddenly increased our knowledge of Magellanic stars, not only
reaching fainter magnitudes but also revealing objects populating the
right center of several stellar clusters (see e.g. Brocato et al. 2001, 
Brocato et al. 2003, Rich et al. 2000). 
 However, the theoretical
scenario concerning the evolution of MC stars has not been
exhaustively explored yet. Recent theoretical computations for
suitable chemical abundances, as given by the``canonical" values
Z=0.004 (SMC) and Z=0.008 (LMC), all rely on the assumption of an
efficient overshooting in the convective cores (Yi et al. 2001,
Girardi et al. 2000, hereinafter GBBC, Salasnich et al. 2000). The
parallel availability of evolutionary results from the canonical
scenario, where convective mixing is based on the well known
Schwarzschild criterion, appears of obvious relevance, at least to
allow significant tests of the various assumptions.

This paper presents such a canonical evolutionary
scenario, as based on the most recent version of our evolutionary
code. For the sake of completeness, evolutionary computations with
efficient overshooting will be also presented, and a comparison with
similar computations in the recent literature will be provided.
The availability of such an evolutionary scenario has  recently allowed
Brocato et al. (2003) to discuss new observational data for the LMC
cluster NGC1866, showing that the question about the actual efficiency
of core overshooting is indeed still open.  In the
following section we will discuss the theoretical background, giving a
preliminary discussion on the choice of several evolutionary inputs
and exploring the robustness of evolutionary results vis-a-vis the
unpredictable amount of mass loss. Evolutionary models and cluster
isochrones are presented in Sect.3, while in Sect.4 models with
and without overshooting are compared with previous results appeared
in the literature. In Sect.5 several parameters of observational
relevance are discussed. A final discussion will close the paper.
\section{The theoretical background}
Stellar models critically depend on the physical
inputs adopted in the computational procedure, and only suitable
observational tests can give light on the actual adequacy of
theoretical predictions. Present models were computed with an updated
version of the FRANEC evolutionary code (see e.g. Chieffi \& Straniero
1989) by adopting recent physical ingredients available in the
literature, namely the equation of state and the opacity from the
Livermore tables (Iglesias \& Rogers 1996, Rogers et al. 1996)
 and  updated nuclear cross sections (see Ciacio et al. 1997 and
 Cassisi et al. 1998 for more details).
When efficient, the element diffusion has been taken into account with
diffusion coefficients as given by Thoul et al. (1994).
With these choices we have already shown that the Standard Solar Model
appears in good agreement with helioseismological constraints (see
Degl'Innocenti et al. 1997) as well as with the location in the
color-magnitude diagram (CMD) of stars with Hipparcos parallaxes in
the two nearby open clusters Hyades and Pleiades (Castellani et al. 2001,
 Castellani et al. 2002).

When moving to the lower metallicities of MC stellar populations,
further tests are needed, to assess the overall adequacy of the
evolutionary predictions and, in particular, to calibrate the value of
the mixing length parameter, which governs the efficiency of the
super-adiabatic convection and, in turn, the temperature of the Red
Giant Branch (RGB) in old stellar clusters. For this test we choose
the well observed galactic globular 47 Tuc with [Fe/H]=-0.70 $\pm$
0.03 (Carretta \& Gratton 1997) and [$\alpha$/Fe]$\approx$0.16 (see
e.g. Ferraro et al. 1999); one indeed derives a total metallicity
Z$\approx$0.004$\div$0.0045, in reasonable agreement with the value
Z=0.004 generally adopted for the SMC population.  As shown in Fig. 1,
by adopting the cluster reddening E(B-V)=0.03 from Schlegel et al.
(1998), our models appear in good agreement with observations, and the
choice l=$\alpha$H${\mathrm p}$ with $\alpha$=1.9 appears able to
reproduce the observed RGB color. The distance modulus (13.56 mag.)
obtained from the fit of the horizontal branch appears in perfect
agreement with the result by Carretta et al. (2000).
\begin{figure*}
   \centering
   \includegraphics[width=8cm]{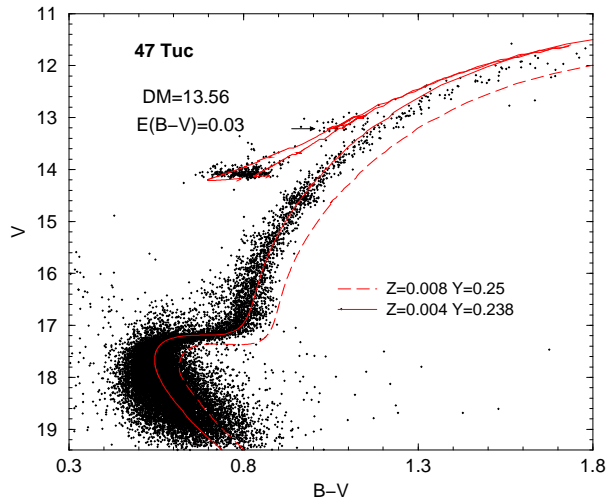}
\caption{The observed CM diagram of 47 Tuc (Sosin et al. 1996) with
superimposed 10 Gyr isochrones with Z=0.004 Y=0.238 (solid line) and
Z=0.008 Y=0.25 (long dashed line), as evaluated by adopting a mixing
lenght parameter $\alpha$=1.9.  Two horizontal branch models,
corresponding to the Z=0.004 isochrone, are also shown, together with
the predicted Asymptotic Giant Branch (AGB)
 bottom luminosity (arrow). Color transformations are
from Castelli (1999).}
\end{figure*}

One may add the encouraging evidence that theoretical predictions
concerning the bottom luminosity of the Asymptotic Giant Branch (AGB) appear in excellent
agreement with observations. Entering into details, one may object
that the predicted RG branch at the larger luminosity becomes
progressively hotter than observed, making the fit good but not
perfect.  One could account for such an evidence by slightly
decreasing the mixing length above the Horizontal Branch (HB) luminosity. However, it
appears worth noting that such a behavior could be explained
 by mass loss in the upper RGB region, with a
final decrease in mass by about 0.1-0.2 M$_{\odot}$. Indeed, if the star
internal structure does not react to mass loss (Castellani \&
Castellani 1993) the stellar envelope moves toward the redder
RG branch of the decreasing masses.

In principle, there are no reasons for the mixing length remaining
constant for stars of different mass or chemical composition.  However,
from the same Fig. 1 one finds that, when passing from Z=0.004 to Z=0.008,
the predicted RGB color increases by about $\Delta$(B-V)$\sim$ 0.1,
in reasonable agreement with the observed difference between SMC and
LMC field Red Giants (Matteucci et al.  2002).  Moreover, Brocato et
al. (2003) have already found that this mixing length calibration
gives red giants in good agreement also with the CM diagram location
of the He burning intermediate mass stars in the LMC cluster NGC1866, with
an estimated metallicity Z=0.007$\div$0.008. As a conclusion, one may
be rather confident that present theoretical results should be at
least in reasonable agreement with the major observational
constraints.

Evolutionary computations have been performed neglecting the
occurrence of mass loss. In the case of low mass stars developing
electron degeneracy in the RGB phase, one knows that mass loss is governing - as a free parameter - the color
location of the Zero Age Horizontal Branch (ZAHB) stars.
 A theoretical analysis of the effect of mass loss on stars
with masses in the range 1.5 M$_{\odot} \div$2.5 M$_{\odot}$ has been
presented by Castellani et al. (2000), showing that one expects only
minor variations in the luminosity of clumping He burning stars.
The effects of mass loss in more massive stars has been already
exhaustively discussed in the literature (see, e.g., Bertelli et al. 1985). 
As an example Fig. 2 shows the evolution of a 4.0
M$_{\odot}$ model as computed by adopting the Reimers's formulation
(Reimers 1975) under various assumptions on the free
parameter $\eta$, governing the mass loss rate. One finds that even for
$\eta$ values as large as 5 (against the recommended value of 1) the major
effect is a reduction of the He burning blue loop, with a decrease in
the loop luminosity not larger than $\Delta$logL$\sim$0.025.

One may conclude that theoretical predictions appear rather robust
vis-a-vis reasonable amounts of mass loss, which should not
sensitively alter the current evolutionary scenario.
\begin{figure*}
\centering
   \includegraphics[width=8cm]{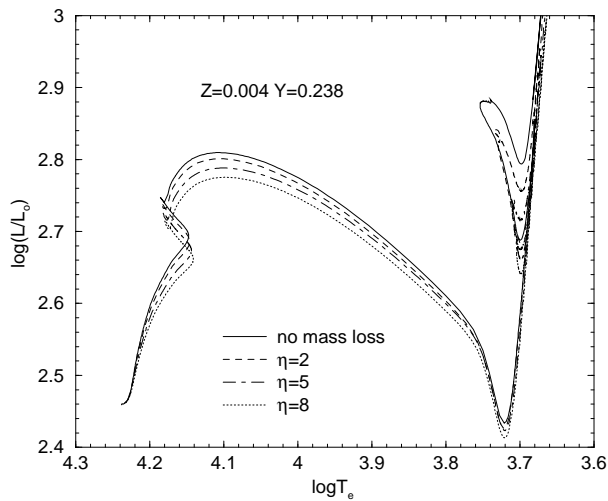}
\caption{The evolutionary path of a 4.0 M$_{\odot}$ model (Z=0.004
Y=0.238) as computed without mass loss ($\eta$=0, solid line) or, top
to bottom, for $\eta$= 2, 5, 8.}
\end{figure*}
\section{The models}
\begin{figure*}
\centering
   \includegraphics[width=8cm]{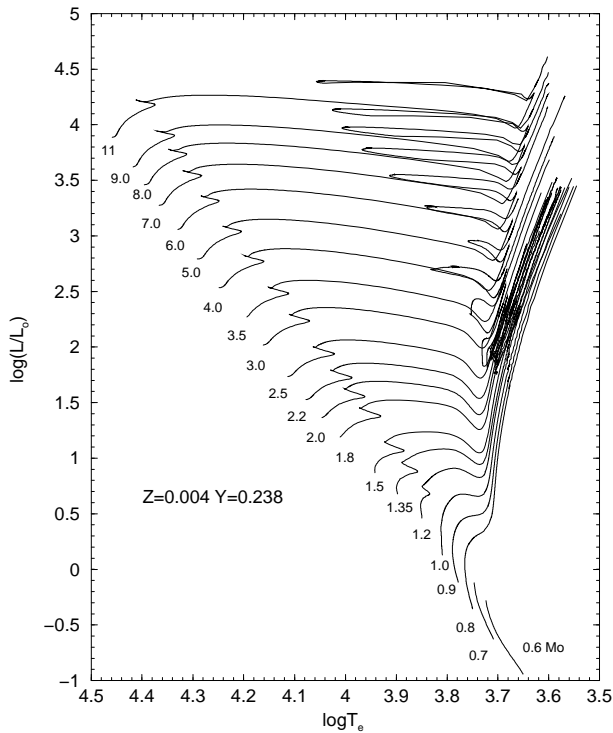}
\caption{Evolutionary path in the log(L/L$_{\odot}$), logT$_{\rm
e} {\rm [^o K]}$ diagram of the models computed for Z=0.004 Y=0.238 and the
labelled mass values.}
\end{figure*}
\begin{figure*}
\centering
   \includegraphics[width=8cm]{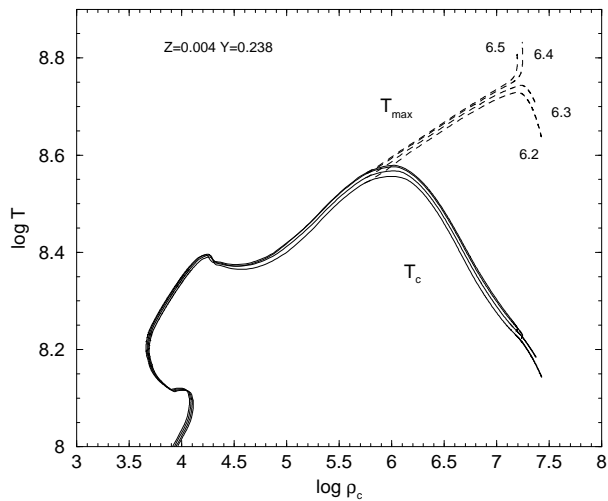}
\caption{The behavior of selected masses close to M$_{\rm up}$ in
the logT${\rm [^o K]}$ versus log$\rho_{\rm c}$ [g/cm$^3$] plane for Z=0.004
Y=0.238. The figure shows the behavior of the central temperature
(solid line) and of the maximum temperature (dashed line) as a function of
 central density.}
\end{figure*}
\begin{figure*}
\centering
   \includegraphics[width=8cm]{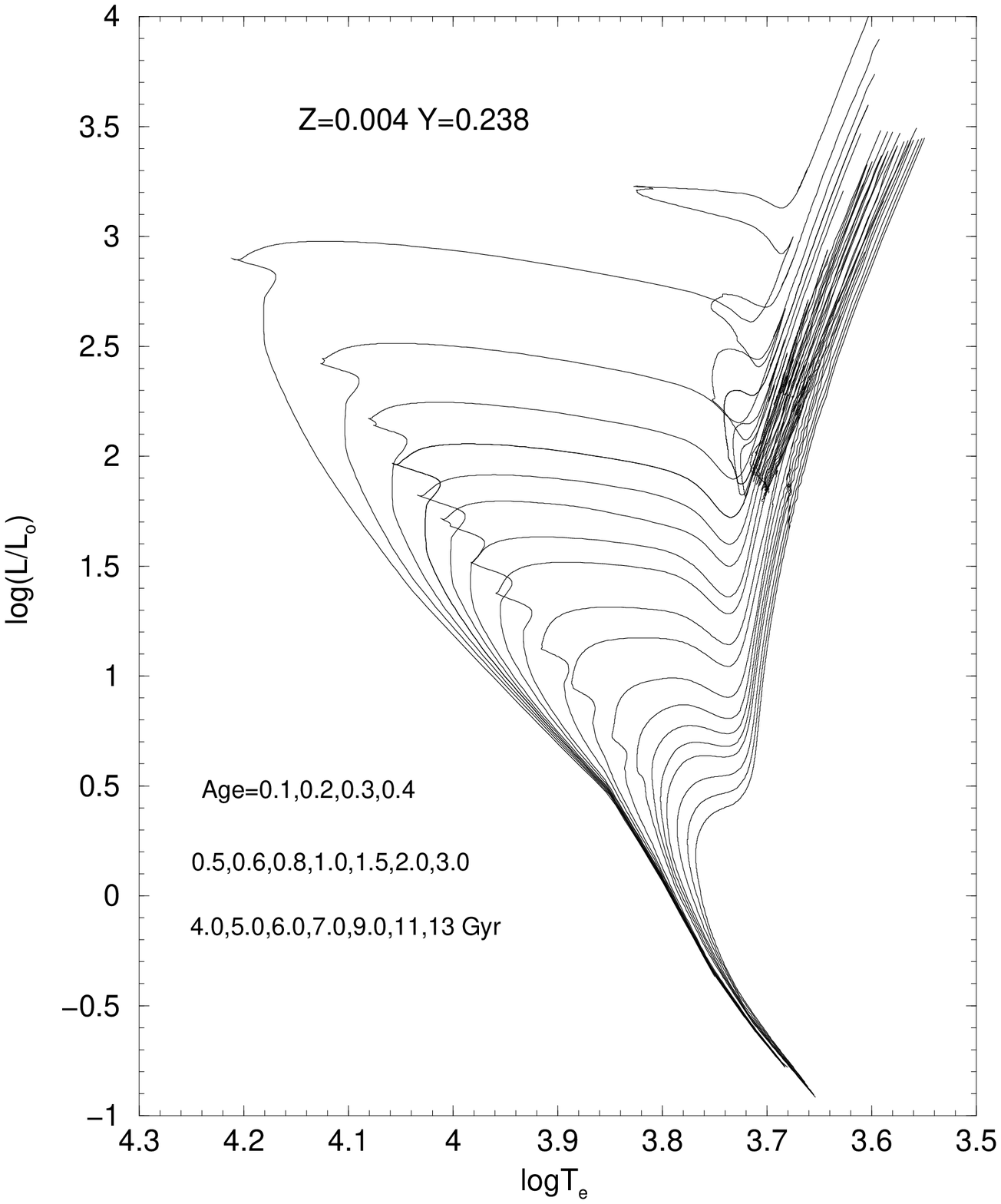}
\caption{Isochrones for Z=0.004 Y=0.238 and the labeled ages.}
\end{figure*}

Suitable sets of evolutionary models covering the mass range 0.8$\div$8
M$_{\odot}$ have been computed for the two metallicities Z=0.008 and
Z=0.004, taken as representative of the Large and the Small Magellanic
Cloud abundances, respectively (see e.g. Luck et al.  1998, Hilker et
al. 1995). Original helium abundances have been fixed at Y=0.238
(Z=0.004) or Y=0.25 (Z=0.008), as obtained by assuming a primordial
helium content Y$_{\rm P}=0.23$ and $\Delta$Y/$\Delta$Z $\sim$2.5 (see
e.g. Pagel \& Portinari 1998, Castellani et al. 1999). 
All these models have been followed from the Main Sequence (MS)
through both the H and He burning phases, till C ignition or the
onset of thermal pulses in the advanced AGB phase. For the less
massive stars undergoing violent He flashes this has been done by
using structures at the RG tip to produce suitable models of ZAHB,
 which have been evolved till the onset of the
thermal pulses. Even less massive
stars (M$\le$0.7~M$_{\odot}$), whose evolutionary times are longer than
the Hubble time, have been evolved up to central H exhaustion.

As an example, Fig. 3 shows the evolutionary paths in the HR diagram
of a set of models for the labelled values of the mass and chemical
composition. Detailed tables for all the tracks are available
at the site http://gipsy.cjb.net in the directory ``Pisa evolutionary
library''.  Each table lists, in the order, the age and, for each
given age, the mass, luminosity and effective temperature, followed by the
visual magnitude and the colors B-V, U-B, V-I, V-R and R-I as derived
by adopting the model atmospheres by Castelli (1999). Tables with the
ZAHB luminosity, temperature, V magnitude and B-V color for the
various masses and  chemical compositions are
also available.  Moreover, in the same site one may find also files
depicting in some details the evolutionary results concerning three
selected masses (M=0.9, 2.0 and 4.0 M$_{\odot}$)
covering the so called RG transition (see e.g. Sweigart et al. 1990).
These tables are intended to
offer to the evolutionary people the opportunity of a close inspection
into our results, allowing significative comparison among different
evolutionary computations. For each given value of the mass, a file
lists in the order the sequence number of the model, its age, central
abundance by mass of H or He, luminosity, effective temperature,
central temperature and density, the maximum off-center temperature
and its location, the mass of the convective core, He core and
convective envelope and, in the last four columns, the fraction of
the total luminosity released by pp, CNO and He nuclear burning and
by the gravothermal energy. The masses of the He core at the helium ignition are in
agreement with the results of Dominguez et al. (1999).

 As a whole,
the evolutionary behavior of all the models follows the well known
prescriptions already and abundantly documented in the literature, and
it does not deserve further comments.  As shown in Fig. 4, we only
notice that for Z=0.004 Y=0.238 one finds the lower mass limit for the
carbon ignition (M$_{\rm up}$) between 6.3 and 6.4 M$_{\odot}$ (for
Z=0.004 Y=0.27 M$_{\rm up}$ is between 6.0 and 6.1 M$_{\odot}$ and
for Z=0.008 Y=0.25 between 6.7 and 6.8 M$_{\odot}$).  Comparison with
previous (canonical) limits as given in Castellani et al. (1990) discloses
 that with the improved input physics the
limit has decreased by about 1 M$_{\odot}$, a result which appears in
agreement with the data presented more recently by Dominguez et
al. (1999). Evolutionary tracks have been finally used to produce
cluster isochrones covering the ages from $\sim$100 Myr to $\sim$15
Gyr.  Figure 5 shows the isochrones set for the case Z=0.004
Y=0.238, as presented in the log(L/L$_{\odot}$), logT$_{\rm e}$
diagram. Data for all the isochrones can be found at the already
quoted web site in files, which, for each given age, give the
mass distribution of the evolving stars and, for each mass, the star
luminosity and effective temperature together with the absolute V magnitude
and selected colors in the Johnson and near-infrared Cousins bands.
\section{Comparison with previous results}
\begin{figure*}
  \centering
   \includegraphics[width=8cm]{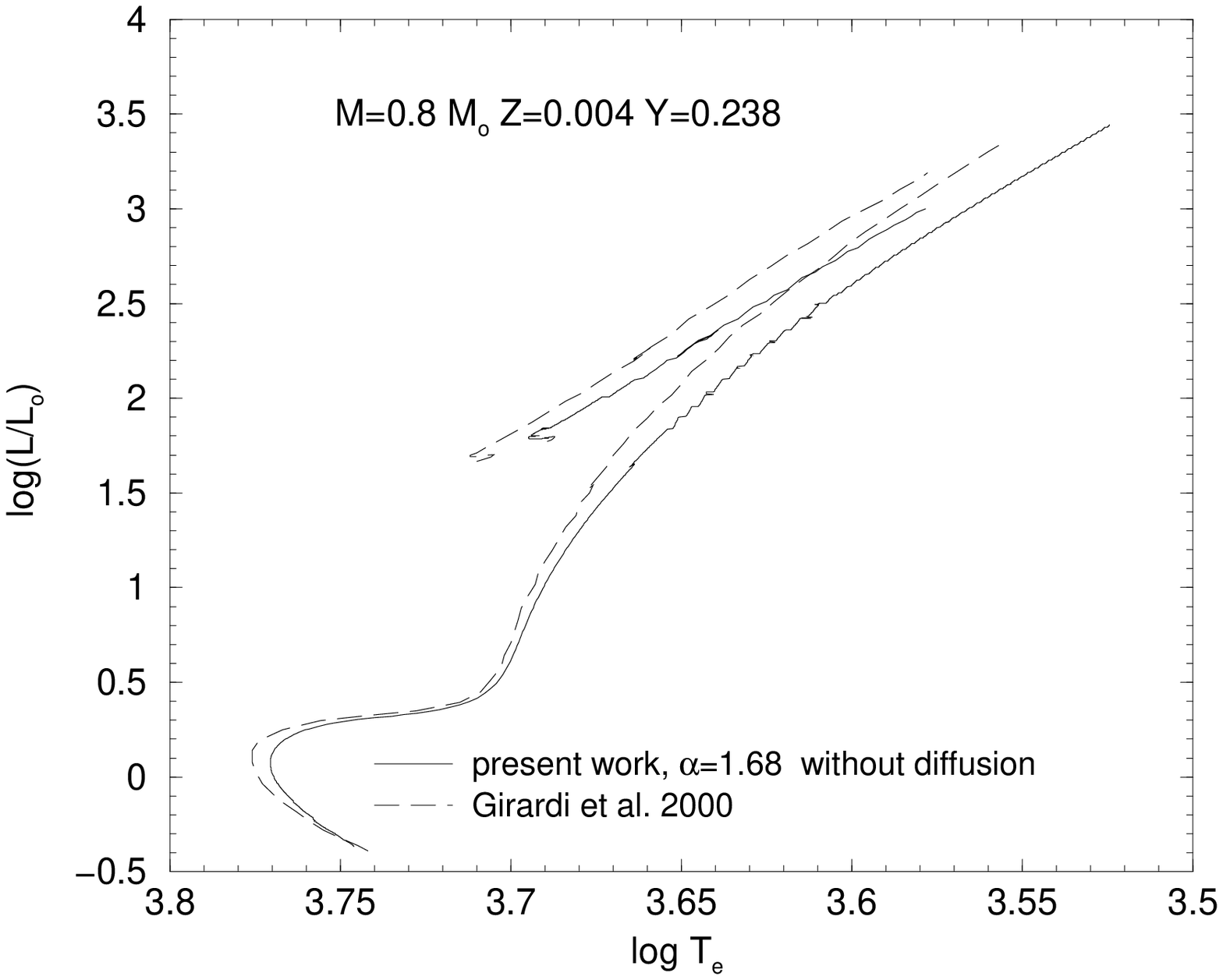}
   \includegraphics[width=8cm]{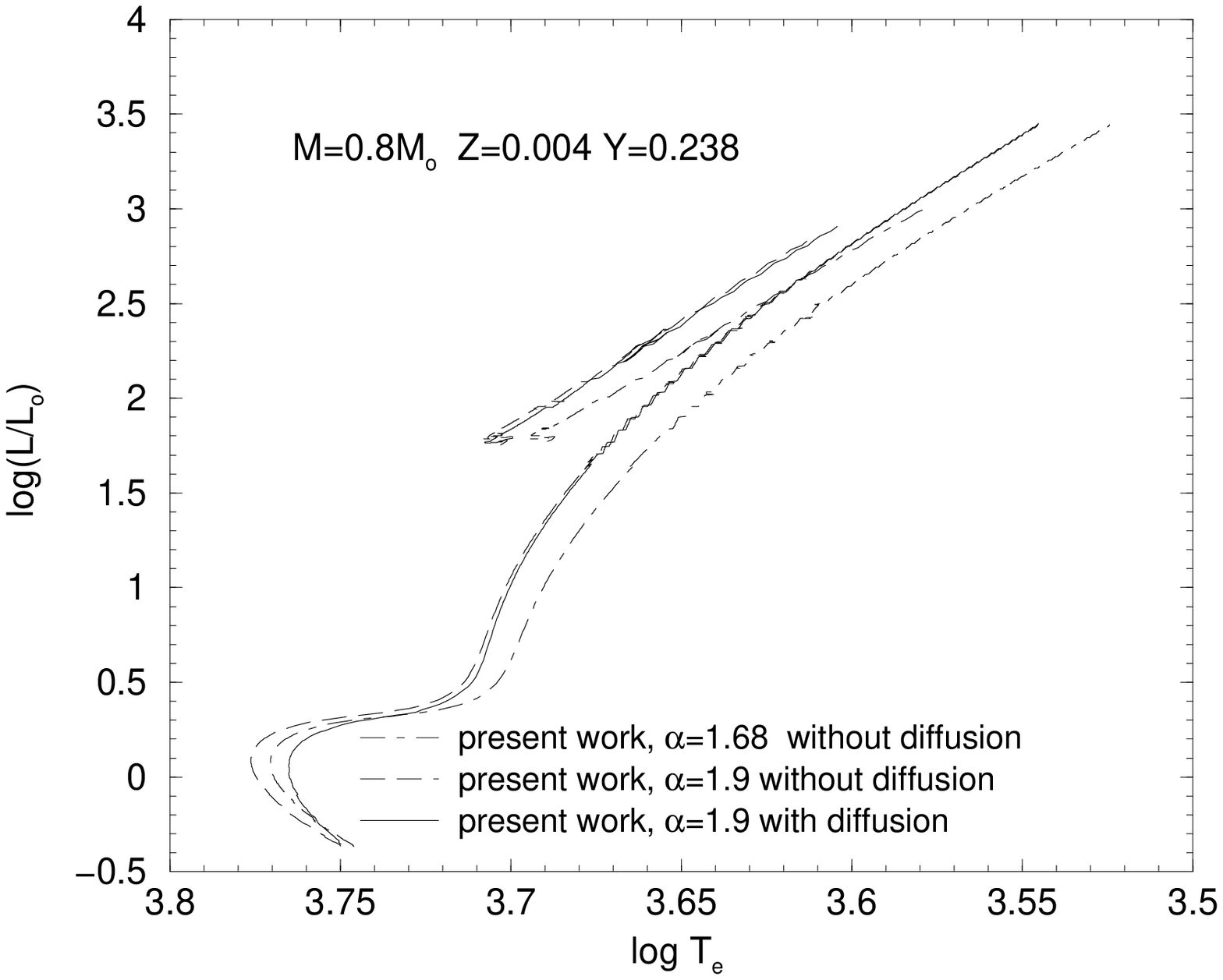}
\caption{Left panel: Comparison
between a 0.8M$_{\odot}$ (Z=0.004 Y=0.24) by Girardi et al. (2000)
and by the present paper with the same characteristics (no
diffusion, $\alpha$=1.68).  Right panel: comparison among our
standard 0.8 M$_{\odot}$ (Z=0.004 Y=0.238) and 0.8M$_{\odot}$ models
with different characteristics (see text).}
\end{figure*}
\begin{figure*}
\centering
   \includegraphics[width=8cm]{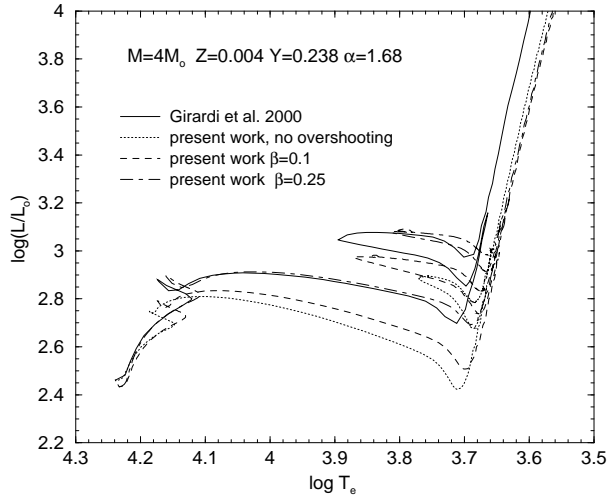}
\caption{Comparison between a 4.0M$_{\odot}$ (Z=0.004 Y=0.238) by
Girardi et al. (2000) (solid line) and by the present paper with the
same characteristics (mild overshooting, $\beta$=0.25, and
$\alpha$=1.68, dot-dashed line line).  Models with $\beta$=0.1 (dashed
line) and without overshooting (dotted line) are also shown.}
\end{figure*}
\begin{table*}
\caption{Comparison among evolutionary models for a star of
0.8M$_{\odot}$ (Z=0.004 Y=0.238) by Girardi et al. (2000) and from the
present work with different characteristics (see text). $\tau_{\rm Xc=0}$
and $\tau_{\rm H_{shell}}$ represent the time (in Gyr) spent in the MS
phase and in the SBG+RGB phases, respectively, while $\tau_{\rm L>1}$
indicates the time in the RGB phase from log(L/L$_{\odot}$)=1 up to
the He flash. LogL$_{\rm ZAHB}$ is the luminosity of the ZAHB
model for the 0.8M$_{\odot}$ mass and $\tau_{\rm HB}$ the time spent in
the central He burning phase.  The value of the R parameter is also
shown.}
\begin{center}
\begin{tabular}{c c c c c c c}
\hline \hline

\\ \hline & $\tau_{\rm Xc=0}$ & $ \tau_{\rm H_{shell}}$ & $\tau_{\rm L>1}$ & $log
L_{\rm ZAHB}$ & $\tau_{\rm HB}$ & R\\

    Girardi et al. 2000 & 15.881 & 2.273 & 0.243 & 1.665 & 0.123 &
        1.89 \\ No diffusion & 14.504 & 3.093 & 0.278 & 1.771 & 0.086
        & 1.46 \\ Standard & 13.853 & 3.067 & 0.289 & 1.754 & 0.084 &
        1.33 \\ \hline \hline
\end{tabular}
\end{center}
\end{table*}
This section will be devoted to a comparison with similar results
appeared in the recent literature. In our feeling, such a comparison
is of some relevance, at least as a warning for the common reader
about the amount of differences that can occur among recent
evolutionary computations and against the uncritical use of
evolutionary results. Differences in the results available in the
literature arise from the adoption of different but still acceptable
physical inputs and from the various assumptions about the efficiency
of some macroscopic mechanisms as the microscopic diffusion or the core
overshooting.  In our knowledge, the only recent papers presenting
evolutionary models and isochrones for the MC chemical composition are
the ones by Girardi et al. (2000, Padova models) and by Yi et
al. (2001, Yonsei-Yale models) both adopting a mild overshooting for H
burning structures. Very recently the Padova models have been compared
with the Yonsei-Yale ones (Gallard et al. 2003), here we will discuss
the comparison of our isochrones with the Girardi et al. ones.  

In the GBBC work all the H burning
models with masses $\ge$1.5 M$_{\odot}$ are supposed to be affected
by the core overshooting, with an extra-mixing extending by about
l$_{\rm ext}=0.5$ H$_{\rm p}$ outside the canonical convective
core. A similar overshooting has been also applied to all convective
cores affecting the central He burning structures.  The convective
envelopes have been computed by adopting a mixing length parameter
$\alpha$=1.68 but allowing for an efficient undershooting at the
bottom of the convective region with l$_{\rm ext}=0.25$
H$_{\rm p}$ for 0.6$\le$M/M$_{\odot}$ $\le$2.0 and l$_{\rm
ext}$=0.7H$_{\rm p}$ for M$>$2.5 M$_{\odot}$. Microscopic
diffusion is not taken into account.

A first meaningful comparison between present and GBBC models can
be made for the H burning phases of low mass stars, where
overshooting is inefficient for the lack of the convective cores. In
Fig.6 (left panel) the GBBC track for a 0.8 M$_{\odot}$ star
(Z=0.004 Y=0.24) is compared with a similar model  as computed
with our code neglecting diffusion and adopting the same chemical
composition and the same mixing length ($\alpha$=1.68). The right
panel of the same figure shows how  evolutionary tracks are varied
when moving toward our canonical model, first increasing  from
$\alpha$=1.68 to 1.9 and, finally, accounting for element
diffusion.  As a whole, differences in the left panel have to be
ascribed to different input physics, as already exhaustively
discussed in Castellani et al. (2000), and to the assumption of
the undershooting in GCCB model. Evolutionary tracks in the right panel
follow well known theoretical predictions (see e.g. Proffit \&
Vandenberg 1991, Chaboyer et al. 1992, Castellani
et al. 1997) for the effects of the element diffusion. As a whole,
the large variations one finds in both panels reinforce the need
for suitable tests and calibrations of theories before producing
evolutionary data for the observational community.
\begin{table*}
\caption{The V magnitude of the MS termination (MT), the difference in
visual magnitude between the clump and MT ($\Delta$M$_{\rm V}$) and the clump mass (m in
M$_{\odot}$) for the isochrones with age $\le$ 6 Gyr (see text) and
chemical composition of LMC (Z=0.008 Y=0.25) are reported for the
``standard'' (without overshooting) case and for the labelled
assumptions on the overshooting parameter $\beta$.}
\begin{center}
\begin{tabular}{c c c c c c c c c c c c}
\hline
\hline
Age & MT & $\Delta$M$_{\rm V}$  & m &  & MT & $\Delta$M$_{\rm V}$  & m &  & MT & $\Delta$M$_{\rm V}$  & m \\
\hline
\multicolumn{1}{c}{}
&\multicolumn{3}{c}{STANDARD}
&\multicolumn{1}{c}{}
&\multicolumn{3}{c}{$\beta$=0.1}
&\multicolumn{1}{c}{}
&\multicolumn{3}{c}{$\beta$=0.25}
\\
\hline
  0.1  & -2.412  &  0.272     &  4.707       &      &  -2.710  &   0.346    &  4.844      &    &    -3.241  &    0.311   &   5.029    \\
  0.2  & -1.363  &  0.564     &  3.516       &      &  -1.627  &   0.550    &  3.595      &    &    -2.050  &    0.414   &   3.731    \\
  0.3  & -0.756  &  0.745     &  2.961       &      &  -1.002  &   0.704    &  3.036      &    &    -1.403  &    0.544   &   3.164    \\
  0.4  & -0.328  &  0.711     &  2.647       &      &  -0.557  &   0.665    &  2.712      &    &    -0.963  &    0.596   &   2.828    \\
  0.5  &  0.005  &  0.550     &  2.437       &      &  -0.210  &   0.634    &  2.491      &    &    -0.593  &    0.561   &   2.594    \\
  0.6  &  0.279  &  0.200     &  2.289       &      &   0.069  &   0.515    &  2.318      &    &    -0.299  &    0.505   &   2.420    \\
  0.7  &  0.510  & -0.100     &  2.171       &      &   0.307  &   0.422    &  2.199      &    &    -0.046  &    0.482   &   2.282    \\
  0.8  &  0.713  & -0.377     &  2.071       &      &   0.517  &   0.078    &  2.110      &    &     0.175  &    0.404   &   2.171    \\
  0.9  &  0.887  & -0.606     &  1.989       &      &   0.701  &  -0.232    &  2.018      &    &     0.374  &    0.242   &   2.080    \\
  1.0  &  1.044  & -0.766     &  1.923       &      &   0.866  &  -0.444    &  1.950      &    &     0.551  &    0.168   &   1.999    \\
  1.1  &  1.184  & -0.910     &  1.867       &      &   1.016  &  -0.582    &  1.891      &    &     0.709  &   -0.036   &   1.947    \\
  1.2  &  1.307  & -1.035     &  1.815       &      &   1.152  &  -0.763    &  1.840      &    &     0.853  &   -0.256   &   1.882    \\
  1.3  &  1.421  & -1.144     &  1.772       &      &   1.277  &  -0.892    &  1.792      &    &     0.974  &   -0.433   &   1.832    \\
  1.4  &  1.527  & -1.242     &  1.732       &      &   1.393  &  -1.015    &  1.750      &    &     1.103  &   -0.597   &   1.791    \\
  1.5  &  1.626  & -1.334     &  1.699       &      &   1.501  &  -1.130    &  1.712      &    &     1.211  &   -0.717   &   1.747    \\
  1.6  &  1.719  & -1.420     &  1.664       &      &   1.602  &  -1.240    &  1.677      &    &     1.315  &   -0.845   &   1.710    \\
  1.7  &  1.806  & -1.501     &  1.635       &      &   1.696  &  -1.340    &  1.644      &    &     1.400  &   -0.941   &   1.676    \\
  1.8  &  1.888  & -1.576     &  1.607       &      &   1.785  &  -1.440    &  1.615      &    &     1.494  &   -1.059   &   1.647    \\
  1.9  &  1.957  & -1.640     &  1.581       &      &   1.868  &  -1.520    &  1.587      &    &     1.576  &   -1.153   &   1.615    \\
  2.0  &  2.013  & -1.690     &  1.557       &      &   1.939  &  -1.580    &  1.561      &    &     1.666  &   -1.256   &   1.587    \\
  2.5  &  2.274  & -1.925     &  1.455       &      &   2.231  &  -1.870    &  1.457      &    &     2.012  &   -1.641   &   1.476    \\
  3.0  &  2.464  & -2.099     &  1.379       &      &   2.455  &  -2.090    &  1.379      &    &     2.251  &   -1.873   &   1.399    \\
  3.5  &  2.664  & -2.273     &  1.314       &      &   2.645  &  -2.244    &  1.317      &    &     2.470  &   -2.077   &   1.336    \\
  4.0  &  2.786  & -2.392     &  1.266       &      &   2.801  &  -2.390    &  1.266      &    &     2.656  &   -2.258   &   1.285    \\
  4.5  &  2.947  & -2.524     &  1.216       &      &   2.939  &  -2.528    &  1.223      &    &     2.823  &   -2.414   &   1.240    \\
  5.0  &  3.021  & -2.601     &  1.185       &      &   3.039  &  -2.61     &  1.186      &    &     2.951  &   -2.535   &   1.202    \\
  5.5  &  3.162  & -2.708     &  1.148       &      &     -    &    -       &    -        &    &      -     &    -       &    -       \\
  6.0  &  3.255  & -2.788     &  1.119       &      &     -    &    -       &    -        &    &      -     &    -       &    -       \\
\hline
\hline
\end{tabular}
\end{center}
\end{table*}
Table 1 compares selected evolutionary quantities for the above
0.8 M$_{\odot}$ models. Even neglecting element diffusion one finds that
the GBBC models have longer central H burning times, but shorter
H-shell lifetimes. The large difference in the central
He burning structures is partially due to the lower luminosity of
their ZAHB models, but we guess it to be largely connected to the
adoption of the core overshooting instead of the canonical semiconvection,
as in our models. Without entering in further details, we feel that
the above comparison convincingly demonstrate that all current
models can be safely adopted to obtain a thumbnail understanding
of the evolutionary status of actual cluster stars. However, before
attempting precise comparison and/or calibrations, one should
preliminarily make a reasoned choice among the various available
models. This is, e.g., shown by the last column in Table 1, which
discloses how far the calibration of the R parameter in terms of
the original He content depends on the adopted evolutionary
scenario.
\begin{table*}
\caption{As in Table 2 but for the metallicity of SMC (Z=0.004) and for the two studied values of
the original helium abundance.}
\begin{center}
\begin{tabular}{c c c c c c c c c c c c}
\hline
\hline
Age & MT & $\Delta$M$_{\rm V}$  & m &  & MT & $\Delta$M$_{\rm V}$  & m & & MT & $\Delta$M$_{\rm V}$  & m
\\
\hline
\multicolumn{1}{c}{}
&\multicolumn{3}{c}{STANDARD Y=0.238}
&\multicolumn{1}{c}{}
&\multicolumn{3}{c}{$\beta$=0.25 Y=0.238}
&\multicolumn{1}{c}{}
&\multicolumn{3}{c}{STANDARD Y=0.27}
\\
\hline
  0.1  &    -2.463 &     0.221 &  4.559      &   &  -3.389  &    0.152   & 5.154    &       &   -2.371 &  0.232 &   4.290  \\
  0.2  &    -1.397 &     0.356 &  3.453      &   &  -2.139  &    0.193   & 3.682    &       &   -1.305 &  0.305 &   3.246  \\
  0.3  &    -0.781 &     0.525 &  2.904      &   &  -1.533  &    0.351   & 3.170    &       &   -0.71  &  0.222 &   2.751  \\
  0.4  &    -0.346 &     0.585 &  2.566      &   &  -1.086  &    0.404   & 2.843    &       &   -0.286 &  0.036 &   2.454  \\
  0.5  &    -0.020 &     0.387 &  2.358      &   &  -0.730  &    0.400   & 2.607    &       &   0.046  & -0.126 &   2.252  \\
  0.6  &    0.273  &     0.115 &  2.204      &   &  -0.437  &    0.385   & 2.432    &       &   0.322  & -0.244 &   2.092  \\
  0.7  &    0.484  &    -0.218 &  2.097      &   &  -0.187  &    0.383   & 2.290    &       &   0.526  & -0.441 &   1.984  \\
  0.8  &    0.651  &    -0.484 &  2.004      &   &  0.032   &    0.375   & 2.174    &       &   0.705  & -0.620 &   1.897  \\
  0.9  &    0.799  &    -0.622 &  1.930      &   &  0.228   &    0.360   & 2.078    &       &   0.866  & -0.781 &   1.822  \\
  1.0  &    0.934  &    -0.749 &  1.869      &   &  0.397   &    0.323   & 2.002    &       &   1.013  & -0.928 &   1.761  \\
  1.1  &    1.058  &    -0.861 &  1.813      &   &  0.544   &    0.142   & 1.937    &       &   1.156  & -1.014 &   1.709  \\
  1.2  &    1.171  &    -0.969 &  1.766      &   &  0.680   &   -0.033   & 1.880    &       &   1.281  & -1.139 &   1.661  \\
  1.3  &    1.276  &    -1.064 &  1.721      &   &  0.805   &   -0.195   & 1.830    &       &   1.397  & -1.253 &   1.618  \\
  1.4  &    1.374  &    -1.154 &  1.682      &   &  0.921   &   -0.342   & 1.786    &       &   1.502  & -1.362 &   1.580  \\
  1.5  &    1.464  &    -1.237 &  1.646      &   &  1.030   &   -0.486   & 1.744    &       &   1.577  & -1.444 &   1.546  \\
  1.6  &    1.549  &    -1.315 &  1.613      &   &  1.132   &   -0.614   & 1.705    &       &   1.666  & -1.534 &   1.514  \\
  1.7  &    1.629  &    -1.390 &  1.582      &   &  1.228   &   -0.741   & 1.671    &       &   1.734  & -1.591 &   1.485  \\
  1.8  &    1.703  &    -1.459 &  1.554      &   &  1.319   &   -0.857   & 1.639    &       &   1.797  & -1.653 &   1.459  \\
  1.9  &    1.773  &    -1.522 &  1.528      &   &  1.405   &   -0.965   & 1.608    &       &   1.865  & -1.716 &   1.435  \\
  2.0  &    1.839  &    -1.586 &  1.504      &   &  1.487   &   -1.074   & 1.581    &       &   1.918  & -1.763 &   1.412  \\
  2.5  &    2.122  &    -1.846 &  1.403      &   &  1.814   &   -1.464   & 1.469    &       &   2.176  & -1.998 &   1.319  \\
  3.0  &    2.334  &    -2.038 &  1.325      &   &  2.066   &   -1.719   & 1.392    &       &   2.378  & -2.176 &   1.246  \\
  3.5  &    2.504  &    -2.193 &  1.263      &   &  2.284   &   -1.938   & 1.330    &       &   2.546  & -2.318 &   1.188  \\
  4.0  &    2.653  &    -2.330 &  1.211      &   &  2.467      &   -2.123   & 1.279 &       &   2.693  & -2.448 &   1.142  \\
  4.5  &    2.783  &    -2.443 &  1.164      &   &  2.644      &   -2.282   & 1.235 &       &   2.821  & -2.560 &   1.104  \\
  5.0  &    2.898  &    -2.542 &  1.123      &   &  2.833      &   -2.487   & 1.195 &       &   2.936  & -2.661 &   1.070  \\
  5.5  &    3.001  &    -2.633 &  1.087      &   &    -        &      -     &  -    &       &   3.039  & -2.751 &   1.041  \\
  6.0  &    3.096  &    -2.715 &  1.050      &   &    -        &      -     &  -    &       &   3.139  & -2.835 &   1.014  \\
\hline
\hline

\end{tabular}
\end{center}
\end{table*}
A short discussion concerning the comparison between present and
GBBC models for the more massive stars has been already presented by
Brocato et al. (2003) for the case of solar metallicity. To extend the
comparison to Magellanic abundances, we computed an additional set of models
(Z=0.004 Y=0.238, Z=0.008 Y=0.25) by allowing an efficient core
overshooting in the H burning phase, as modelled according to
Castellani et al. (2000) with  ${l}_{\rm ov}=\beta{\rm H}_{\rm p}$, and
$\beta$ values covering the range suggested in the recent
literature (see e.g. Girardi et al. 2000, Pols et al. 1998).
We recall that due to the different treatment of the overshooting phenomenon
the $\beta$ value adopted by the Padua group is roughly equivalent to one
half of the $\beta$ value adopted in present work. In particular  the Padua
 $\beta$=0.5 corresponds to the
present $\beta$=0.25.

\begin{table*}
\caption{The TO visual magnitudes, as a function of the age, for the
studied chemical compositions.}
\begin{center}
\begin{tabular}{c c c c}
\hline
\hline
Age(Gyr) & M$_{\rm V}^{\rm TO}$ & M$_{\rm V}^{\rm TO}$ & M$_{\rm V}^{\rm TO}$
\\
\hline
\multicolumn{1}{c}{}
&\multicolumn{1}{c}{Z=0.008 Y=0.25}
&\multicolumn{1}{c}{Z=0.004 Y=0.238}
&\multicolumn{1}{c}{Z=0.004 Y=0.27}
\\
\hline
7.0      &    3.906  &   3.797  &   3.816  \\
8.0      &    4.032  &   3.919  &   3.922  \\
9.0      &    4.137  &   4.026  &   4.063  \\
10.0     &    4.200  &   4.106  &   4.167  \\
11.0     &    4.232  &   4.196  &   4.268  \\
12.0     &    4.332  &   4.280  &   4.369  \\
\hline
\hline
\end{tabular}
\end{center}
\end{table*}
\begin{table*}
\caption{ZAHB visual magnitude in the RR Lyrae region ($M_{\rm V}^{\rm 3.83}$). The progenitor
mass is 0.8 M$_{\odot}$ for Z=0.004 Y=0.27 and 0.9 M$_{\odot}$ in the
other two cases.}
\begin{center}
\begin{tabular}{c c c}
\hline
\hline
 M$_{\rm V}^{\rm 3.83}$ &  M$_{\rm V}^{\rm 3.83}$ &  M$_{\rm V}^{\rm 3.83}$
\\
\hline
\multicolumn{1}{c}{Z=0.008 Y=0.25}
&\multicolumn{1}{c}{Z=0.004 Y=0.238}
&\multicolumn{1}{c}{Z=0.004 Y=0.27}
\\
\hline
0.761 & 0.672 & 0.565 \\
\hline
\hline
\end{tabular}
\end{center}
\end{table*}
Figure 7 shows the comparison between a 4 M$_{\odot}$ by Girardi et
al. (2000) and our model with a similar amount of overshooting and the
same mixing lenght. One finds that the two models are in rather good
agreement, but with GBBC He burning giants fainter than ours. The same figure
 shows the dramatic difference when neglecting
the overshooting, in agreement with already well known predictions (see
e.g. Maeder 1975, Bressan et al. 1993 and references therein). Data in
the figure add the evidence that a mild overshooting, as given by
$\beta$=0.1, produces only marginal variations in the predicted
evolutionary path, so that one can easily predict that the occurrence
of such a mild overshooting will be very hard to proof. Evolutionary
lines and cluster isochrones with selected choices for the efficiency
of overshooting can be found at the same web link quoted above.  As
usual, models with overshooting are calculated for masses $\ge$1.2 M$_{\odot}$
to avoid the unobserved presence of the overall
contraction feature in low mass stars; for this reason isochrones with
overshooting are calculated for ages lower than about 5 Gyr. Due to
the relatively low ages, models with overshooting do not include microscopic
diffusion.
\section{An age indicator}
\begin{figure*}
\centering
   \includegraphics[width=8cm]{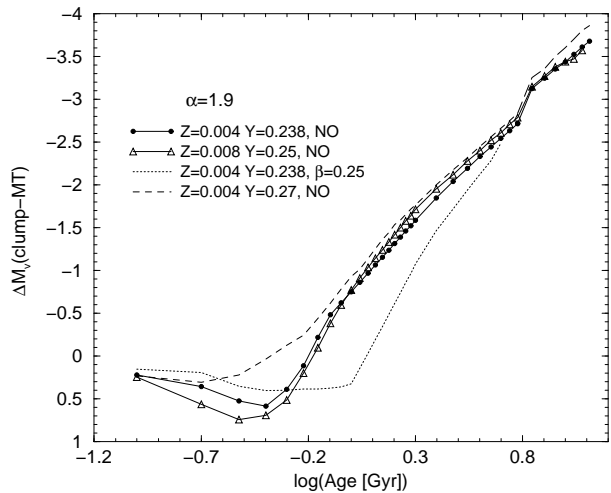}
\caption{The $\Delta$M$_{\rm V}$ parameter  as a function of
the cluster age for the labelled assumptions about the original
chemical composition and the efficiency of core overshooting
(NO= canonical models).  For ages larger than 5 Gyr convective
cores vanish  and  the usual luminosity difference between
HB and TO has been reported.}
\end{figure*}
Theoretical constraints on the age of  stellar clusters are among
the most relevant results of stellar evolutionary theories,
allowing to shed light on the history of stellar populations. In
this context, clusters belonging to the Magellanic Clouds appear
of particular interest, since their significant range of ages
represents an ideal target for testing theory. The most
exhaustive approach to the problem of the cluster ages is
obviously provided by the well known isochrone fitting procedure.
However, the availability of more direct and simple ``age
indicators" has been already proved of some relevance.

In the case of old globulars, since the pioneering paper by Iben
(1968),  the so called ``vertical method'', based on the
difference ($\Delta$M$_{\rm V}$) between the Turn-Off (TO) and the
helium burning HB phase has been widely used (see e.g. Stetson et al. 1996,
 Salaris \& Weiss 1997, Cassisi et al.
1998, 1999). A similar method can be however adopted for younger
clusters, calibrating the difference in magnitude between the
bright MS termination (MT) and the clump of He burning giants
(see e.g. Salaris \& Girardi 2002, Castellani et al. 1999,
Udalski et al. 1998). According
to such an evidence, we made use of present isochrones to
calibrate such a parameter in terms of the cluster ages. The aim
is to provide an easy-to-do observational parameter able to derive
from the CM diagram of a cluster, at least, a correct order of
magnitude for the age, independently of the cluster distance.

Table 2 gives data for such a calibration, as evaluated for LMC
compositions and with the labelled assumptions about the
efficiency of overshooting, whereas Table 3 gives similar data but
for SMC, testing also the effects of an increased amount of
original He. The tables gives, for each value of the cluster age,
the V magnitude of the MS termination and the difference in
visual magnitude between the clump and MT evaluated respectively as 
the bottom luminosity of the He clump region and as 
the brightest magnitude reached just after the overall contraction phase (H exhaustion). 
For each age, the original mass of stars populating the He burning clump is
also reported.

Figure 8 shows the run of $\Delta$M$_{\rm V}$ as a function of the cluster
age. As already known (see, e.g., Fig. 11 in Castellani et al. 1992)
 one finds that $\Delta$M$_{\rm V}$ attains a minimum
for ages around 300-400 Myr, definitely increasing when the age
increases above  that value. One concludes that $\Delta$M$_{\rm V}$
works as an univocal indicator only for ages larger than 700-800
Myr, whereas for lower ages there is in principle a possible
ambiguity. Data in the same figure shows the not negligible
dependence of  the calibration  on the assumptions made about both
overshooting and He content, representing an obvious additional
uncertainty. However, it could offers the opportunity to constrain
these parameters by comparison with a  set of well observed LMC
clusters covering a suitable range of ages.

Interestingly enough, one may notice that the predicted evolutionary
scenario appears at least qualitatively supported by the the
sample of 21 CM diagrams for LMC clusters recently presented by
Brocato et al. (2001). By looking at Fig. 14 in that
paper, one finds (apart from the case
of NGC6718) a support to the overall behavior of $\Delta$M$_{\rm V}$ with
time, including the occurrence of a minimum value of the order of
the  predicted one. Data in the previous Tables 2 and 3 can be
useful in several ways. As an example, by simply looking at the CM
diagram of  cluster NGC2420 (Anthony-Twarog et al. 1990) from the
observed difference in magnitude between the He-clump and the MS
termination ($\Delta$M$_{\rm V}\sim$ 1.5 mag ) one finds for the cluster
a (canonical) age of the order of 2 Gyr, in agreement with the
results obtained through the fit of the CMD diagram (see e.g. Pols
et al. 1998, Prada Moroni et al. 2001) but without the need for a
complicate isochrone fitting procedure. In the meantime one gives,
at least, an indication of the absolute magnitude of the MS
termination and, thus, of the cluster distance modulus together
with information on the original mass of the evolving He giants.
\section{Final remarks}
In this paper we have presented and discussed the canonical
evolutionary scenario for stars with the chemical compositions suitable
for the two MCs. The effect of overshooting has been also discussed
by producing a set of parallel investigations as based on the same
procedures and the same input physics. On this basis we make available
evolutionary tracks and cluster isochrones covering the range of ages
100 Myr$\div$15 Gyr. The difference in magnitudes between the top MS
(the Blue Sequence) and He burning structures has been finally
calibrated in terms of the cluster age.  However, we are well aware that
observational constraints on such a parameter can be sometime not
easy, if not difficult. A poorly populated cluster may lack of the
most luminous stars, either in the He burning phase or above the
overall contraction phase.  Moreover, in strongly populated clusters
the occurrence of binaries can mask the MS top, adding objects at
higher luminosity.  However, even bearing in mind these warnings, we
feel that such a calibration appears as a relevant topic of the
presented evolutionary scenario, allowing an easy and quick approach
to the problem of cluster ages.

\begin{acknowledgements}
Financial support for this work was provided by the Ministero
dell'Istruzione, dell'Universit\`a e della Ricerca (MIUR) under
the scientific project ``Stellar observables of cosmological
relevance'' (V. Castellani \& A. Tornamb\`e, coordinators).
\end{acknowledgements}


\begin{thebibliography}{99}
\bibitem{} Anthony-Twarog B.J., Twarog B.A., Kaluzny J., \& Shara M.M. 1990, AJ 99, 1504
\bibitem{} Bertelli G., Bressan A.G., \& Chiosi C. 1985, A\&A 150,33
\bibitem{} Bertelli G., Nasi E., Girardi L. et al., 2003, ApJ 125, 770
\bibitem{} Bressan A., Fagotto F., Bertelli G., \& Chiosi C. 1993, A\&AS 100, 647
\bibitem{} Brocato E., Castellani V., Raimondo G., \& Walker A. 2003, AJ (in publication), astro/ph 0302458
\bibitem{} Brocato E., Di Carlo E., \& Menna G. 2001, A\&A 374, 523
\bibitem{} Carretta E. \& Gratton R. 1997, A\&AS 121, 95
\bibitem{} Carretta E., Gratton R., Clementini G., \& Fusi Pecci F. 2000, ApJ 533, 215
\bibitem{} Cassisi S., Castellani V., Degl'Innocenti S., \& Weiss A. 1998, A\&AS 129, 267
\bibitem{} Cassisi S., Castellani V., Degl'Innocenti S., Salaris M., \& Weiss A. 1999, A\&AS 134, 103
\bibitem{} Castellani M., \& Castellani V. 1993, Apj 407, 649
\bibitem{} Castellani V., Chieffi A., \& Straniero O. 1990, ApJS 74,463
\bibitem{} Castellani V., Chieffi A., \& Straniero O. 1992, ApJS 78,517
\bibitem{} Castellani V., Ciacio F., Degl'Innocenti S., \& Fiorentini G. 1997, A\&A 322, 801
\bibitem{} Castellani V., Degl'Innocenti S., \& Marconi M. 1999, MNRAS 303, 265
\bibitem{} Castellani V., Degl'Innocenti S., Girardi L. et al. 2000, A\&A, 354, 150
\bibitem{} Castellani V., Degl'Innocenti S., \& Prada Moroni P.G. 2001, MNRAS 320, 66
\bibitem{} Castellani V., Degl'Innocenti S., Prada Moroni P.G., \& Tordiglione V. 2002, MNRAS 334, 193
\bibitem{} Castelli F. 1999, A\&A, 346, 564
\bibitem{} Chaboyer B., Sarajedini A., \& Demarque P. 1992, ApJ 394, 515
\bibitem{} Chieffi A., \& Straniero O. 1989, ApJS 71, 47
\bibitem{} Ciacio F., Degl'Innocenti S., \& Ricci B. 1997, A\&AS 123, 449
\bibitem{} Degl'Innocenti S., Dziembowski W.A, Fiorentini G., \& Ricci B. 1997, Astrop. Phys. 7, 77
\bibitem{} Dominguez I., Chieffi A., Limongi M., \& Straniero O. 1999, ApJ 524, 226
\bibitem{} Ferraro F.R., Messineo M., Fusi Pecci F. et al. 1999, AJ 118, 1738
\bibitem{} Gallart C., Zoccali M., Bertelli G. et al., 2003, ApJ 125, 742
\bibitem{} Girardi L., Bressan A., Bertelli G., \& Chiosi C. 2000, A\&AS 141, 371
\bibitem{} Hilker M., Richtler T., \& Gieren W. 1995, A\&A 294, 648
\bibitem{} Iben I.Jr. 1968, Nature, 220, 143
\bibitem{} Iglesias C.A., \& Rogers F.J. 1996, ApJ 464, 943
\bibitem{} Luck R.E., Moffet T.J., Barnes T.G., \& Gieren W.P. 1998, AJ 115, 605
\bibitem{} Maeder A. 1975, A\&A 40, 303
\bibitem{} Matteucci A., Ripepi V., Brocato E., \& Castellani V. 2002, A\&A 387, 861
\bibitem{} Pagel B.E.J., \& Portinari L. 1998, MNRAS 298, 747
\bibitem{} Pols O.R., Schroeder K-P, Hurley J.R., Tout C.A., \& Eggleton P.P. 1998, MNRAS 298, 525
\bibitem{} Prada Moroni P.G., Castellani V., Degl'Innocenti S., \& Marconi M. 2001, Mem. SAIt 72, 407
\bibitem{} Proffit C.R. \& VandenBerg D.A. 1991, ApJS 77, 473
\bibitem{} Reimers D. 1975, Mem. Soc. R. Sci. Liege, ser. 6, vol.8, p. 369
\bibitem{} Rich R. M., Shara M., Fall S. M., \& Zurek D., 2000, AJ 119, 197
\bibitem{} Rogers F.J., Swenson F.J., \& Iglesias C.A. 1996, ApJ 456, 902
\bibitem{} Salaris M, \& Girardi L., 2002, MNRAS 337, 332 (2002)
\bibitem{} Salaris M, \& Weiss A. 1997, A\&A 327, 107
\bibitem{} Salaris M, Percival S., Brocato E., Raimondo G., \& Walker A. R., 2003, ApJ (accepted), astro-ph/0301532 
\bibitem{} Salasnich B., Girardi L., Weiss A., \& Chiosi C. 2000, A\&A 361, 1023
\bibitem{} Schlegel D.J., Finkbeiner D.P., \& Davis M. 1998, ApJ, 500, 525
\bibitem{} Sosin C., Piotto G., Djorgovski S.G. et al., 1996, Proceeding of the workshop "Stellar Ecology",
Marciana Marina, Italy, Rood R.D. and Renzini A. eds., p.92
\bibitem{} Stetson P.B., VandenBerg D.A., \& Bolte M., 1996, PASP 108, 560
\bibitem{} Sweigart A.V., Greggio L., \& Renzini A., 1990, ApJ 364, 527
\bibitem{} Thoul A., Bahcall J., \& Loeb A. 1994, ApJ 421, 828
\bibitem{} Udalski A., Szyma\'nski M., Kubiak M. et al., 1998, Acta Astr. 48, 1
\bibitem{} VandenBerg D.A., Bolte M., \& Stetson P.B. 1996, ARA\&A 34, 461
\bibitem{} Yi S., Demarque P., Kim Y-C. et al., 2001, ApJ 533, 670
\bibitem{} Woo J. H., Gallart C., Demarque P., Yi S., \& Zoccali M., 2003, ApJ 125, 754
\end{thebibliography}
\end{document}